\documentclass{iaus}
\usepackage{graphicx}

\title[S 265.~~The Galactic bulge] %% give here short title %%
{The Stellar Population of the Galactic Bulge}

\author[Manuela Zoccali]  %% give here short author list %%
{M. Zoccali$^1$}

\affiliation{$^1$Pontificia Universidad Cat\'olica de Chile \\
Casilla 306, Santiago 22, Chile \\ email: {\tt mzoccali@astro.puc.cl}}

\pubyear{2009}
\volume{265}  %% insert here IAU Symposium No.
\pagerange{119--126}
\jname{Chemical Abundances in the Universe: Connecting First Stars to Planets}
\editors{K. Cunha, M. Spite \& B. Barbuy, eds.}
\begin{document}

\maketitle

\begin{abstract}
The Galactic bulge  is the central spheroid of  our Galaxy, containing
about one quarter of the total  stellar mass of the Milky Way (M$_{\rm
bulge}=1.8   \times  10^{10}   M_\odot$;  Sofue,   Honma   \&  Omodaka
2009). Being older than the disk, it is the first massive component of
the Galaxy to have collapsed into stars.  Understanding its structure,
and the  properties of its  stellar population, is therefore  of great
relevance  for galaxy  formation models.   I will  review  our current
knowledge of  the bulge properties, with special  emphasis on chemical
abundances,   recently    measured   for   several    hundred   stars.
\end{abstract}

\firstsection % if your document starts with a section,
              % remove some space above using this command.
\section{The bulge structure}

The near infrared images from the COBE/DIRBE experiment clearly showed
that our Galaxy has a boxy  shaped bulge (Dwek et al.  1995). Isophote
deprojection revealed  a barlike nature, with  axes ratios 1:0.33:0.23
and  with  the  near side  on  the  $1^{\rm  st}$ quadrant,  at  $\sim
20^\circ$ from the Sun-Galactic center direction.  These findings were
later confirmed  by several authors (e.g., Babusiaux  \& Gilmore 2005;
Rattenbury et  al.  2007a, and references therein),  hence the prolate
nature of  the bulge is now  widely accepted. The scale  length of the
{\it main} bar is $\sim 1.5$ kpc.  A smaller bar (scale $\sim 600$ pc)
seems to be also present in  the inner bulge (Alard 2001, Nishiyama et
al.   2005),  though  further   studies  are  needed  to  confirm  and
characterize this structure.

A  striking feature recently  discovered in  the outer  bulge suggests
that the Galactic  bulge is all but a  simple prolate spheroid.  Along
the minor axis, at distances in  excess of $\sim 700$ pc, a double red
clump  is  clearly  visible  in  several  independent  sets  of  data,
symmetric  both  at  positive  and negative  latitudes  (McWilliam  \&
Zoccali  2009). The double  clump disappears  outside the  minor axis,
leaving only the  brighter of the two at  positive longitudes, and the
fainter of  the two at negative longitudes.   Photometric data mapping
the  whole bulge  area are  presently  available only  from the  2MASS
survey,  which  is  not  faint  enough  to reach  the  red  clump  for
$|b|<3^\circ$.  The  {\it Vista Variable  in the Via  L\'actea} survey
(Minniti et al. 2009) will solve this problem, mapping the whole bulge
to much fainter magnitudes ($K_s  \sim 20$ in the coadded images), and
allowing  to deproject  its  3D structure  by  means of  its RR  Lyrae
variable stars.

\section{The bulge age}

According  to galaxy  formation models,  a bar  can be  formed through
secular evolution of the disk. Dynamical instabilities would cause the
bar to  buckle (bend),  then thicken and  eventually look like  a {\it
pseudobulge} (Combes \& Sanders 1981, Combes et al. 1990, Athanassoula
\& Misiriotis  2002, Athanassoula 2005).  Deriving from  disk stars, a
{\it pseudobulge}  would have several properties,  such as kinematics,
stellar  ages, chemical  abundances, resembling  more those  of galaxy
disks rather than those  of classical spheroids (Kormendy \& Kennicutt
2004).

\begin{figure}[t]
\begin{center}
\includegraphics[width=3.4in]{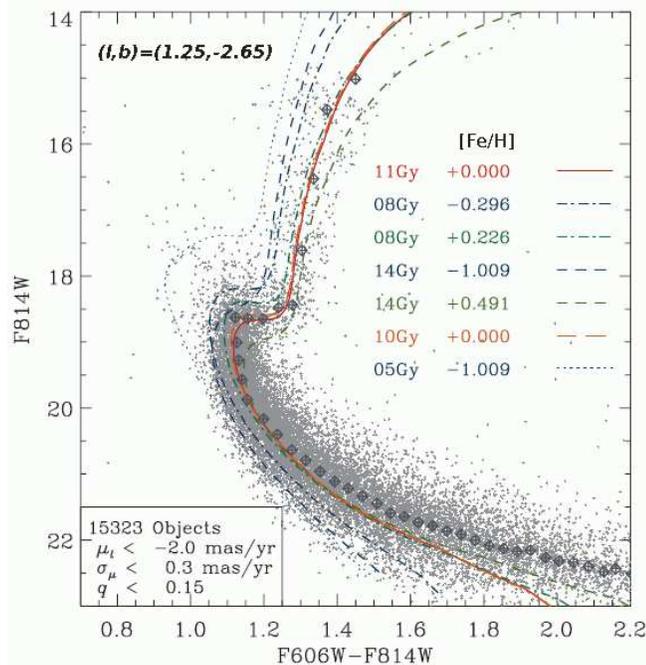}
\caption{The CMD of a clean sample of bulge stars, selected on the 
basis of their proper motions. Also shown are isochrones for different
ages and metallicities. It is evident that bulge stars follow the old 
isochrones, with the precise age depending on the adopted metallicity.
The residual stars brighter than the old turnoff do not follow a younger
isochrone, and indeed Clarkson et al. (2009) argues that they are likely
to be all blue stragglers. Figure adapted from Clarkson et al. (2008).}
   \label{cmd}
\end{center}
\end{figure}

In contrast with that, Ortolani et al.  (1995) first demonstrated that
the   stellar  population  of   the  bulge,   in  Baade's   Window  at
$(l,b)=(0,-4)$,  is  as old  as  the  stars  in the  globular  cluster
47~Tucanae.  This  result was later  confirmed by Feltzing  \& Gilmore
(2000),  Kuijken \&  Rich (2002),  Zoccali  et al.   (2003), and  more
recently by  Clarkson et al. (2008)  who excluded disk  stars from the
color magnitude  diagram (CMD)  on the basis  of their  proper motions
(Fig~\ref{cmd}).   It should  be noted,  however, that  all  the above
works studied the bulge stellar  population in small fields very close
to the  minor axis.   Dynamical simulation of  bulge formation  in the
bar-driven scenario suggest that the vertical heating is significantly
larger at the  two ends of the bar, indicating  these positions as the
places where  the intermediate age  components should be  found (e.g.,
Debattista et al.   2004; and references therein).  Due  to the larger
interstellar  extinction, deep photometry  reaching the  main sequence
turnoff  away  from  the  minor  axis  is not  yet  available  in  the
literature.  Some  data, acquired with  the near IR  camera HAWKI@VLT,
are  under analysis  (Valenti  et  al.  2010).  Brown  et al.   (2009)
present a  new photometric system  employing five WFC3  bands spanning
the  UV,  optical, and  near  IR,  especially  designed to  break  the
degeneracy  between reddening,  temperature and  metallicity,  for the
forthcoming  WF3 Galactic  bulge Treasury  Program. This  program will
allow us  to derive the bulge  star formation history  in four fields,
including one at the far edge of the bar.

\section{The bulge metallicity}

The mean age of the stars,  as measured from the magnitude of the main
sequence turnoff, is  the most direct tool to date  the formation of a
stellar system.  However, the  sensitivity of the turnoff magnitude to
age decreases  for older  stellar populations. The  difference between
two solar metallicity isochrones of 10 and 12 Gyr, respectively, is of
$\Delta  M_V  \approx 0.18$  magnitudes,  thus  requiring a  precision
currently  impossible to obtain  for stars  in the  bulge, due  to the
instrinsic spread in distance, metallicity and differential reddening,
all concurring to smear out the features of the CMD.

Luckily enough, the chemical composition of the stars has the opposite
behaviour.   Significant changes  in the  chemical composition  of the
interstellar medium,  hence on the  chemical composition of  the newly
born stars,  occur during  the first $\sim  1$ Gyr from  the formation
epoch of a stellar system.  During that time, indeed, massive stars of
different masses explode as  core collapse supernovae (SNe), and later
on the first thermonuclear SNe  start exploding too, all enriching the
medium of different kind of elements.

The bulge metallicity distribution function (MDF) was first determined
by McWilliam  and Rich (1994) who obtained  high resolution (R=17,000)
spectra of  11 bulge stars, and  used them to calibrate  a larger (88)
sample of low resolution spectra  from Rich (1988). They found a broad
MDF, peaked at solar metallicity, with a shape roughly compatible with
that of a {\it closed box} model.  Mentioning here only works based on
high resolution spectroscopy, a new  determination of the MDF was made
by Fulbright  et al.   (2006) using spectra  for 27 stars  observed at
R$\sim$60,000 to  re-calibrate a sample  of 217 stars observed  at low
resolution  by Sadler  et  al.   (1996).  The  resulting  MDF is  very
similar, only much  smoothed, to the one by  McWilliam \& Rich (1994).
Both these studies were confined to the low extinction Baade's Window.
Investigating  the presence of  a radial  metallicity gradient  in the
bulge obviously requires the  observations of many fields, which until
a few years ago was only possible through photometry or low resolution
spectroscopy.   Minniti  et al.   (1995),  using  both  their own  and
literature  data, claimed  the  presence of  a  radial gradient,  from
[Fe/H]$\sim  +0.2$   in  Baade's   Window  ($\sim  600$pc),   down  to
[Fe/H]$\sim -1$  at 2.3 kpc from  the Galatic center,  along the minor
axis. Ramirez et  al. (2000) and Rich et  al. (2007a), however, claimed
the absence  of a  gradient from Baade's  Window inward.  It  is worth
noticing  that the  work  by Rich  et  al.  (2007a)  is  based on  high
resolution near IR spectra, for a sample of $\sim 15$ stars in each of
two fields at $140$ and $600$ pc, respectively.

\begin{figure}[t]
% \vspace*{-2.0 cm}
\begin{center}
 \includegraphics[width=3.4in]{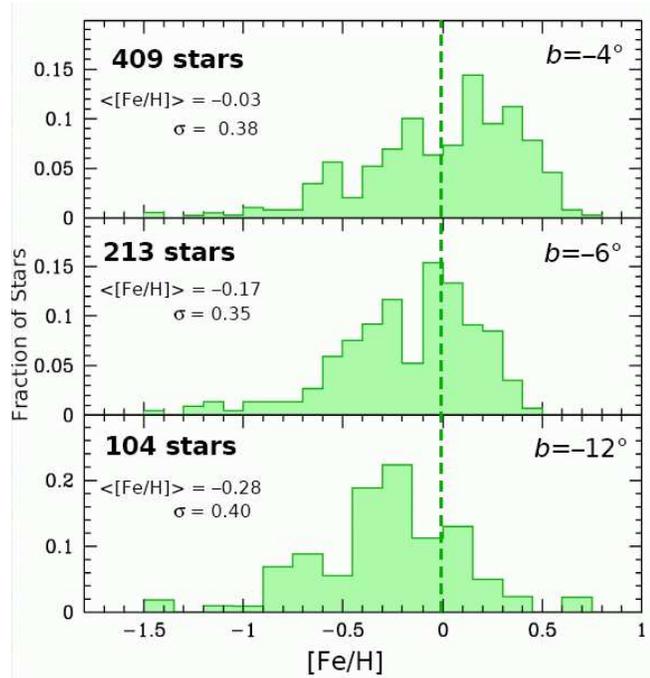} 
% \vspace*{-1.0 cm}
 \caption{The  bulge MDF  in three  fields  along the  minor axis,  at
latitudes  listed on the  top right  corners.  Mean  metallicities and
dispersions are labeled.}
   \label{mdf}
\end{center}
\end{figure}

The advent of  the multifibre spectrograph FLAMES allowed  a huge step
forward in the  field. Zoccali et al. (2008)  acquired GIRAFFE spectra
for  $\sim  700$  bulge  K  giant  stars  in  four  different  fields,
containing 3 globular clusters.  Spectra for about 200 red clump stars
in  Baade's Window, observed  through the  same setups and conditions,
were added to the sample,  from the french GIRAFFE GTO programme (Hill
et al. 2009)  The analysis of the stars in the  fields along the bulge
minor axis allowed to determine the MDF shown in Fig.~\ref{mdf}, which
is   the  first  one   derived  from   all  high   resolution  spectra
(R=20,000). Clearly,  the mean metallicity is higher  in the innermost
field, and  decreases towards  the outer one.  It might also  be noted
that  rather than  a solid  shift of  the MDF  towards the  metal poor
regime,  going outwards  the metal  rich  side becomes  less and  less
populated, in favor  of the metal poor one,  thus suggesting something
like  a different  proportion of  populations at  different  radii. It
should be emphasized that all three samples suffer from some degree of
contamination from  disk stars.  The amount of  disk contamination, at
present,  can only be  {\it estimated},  e.g., using  the Besan\c{c}on
Galaxy model (Robin  et al.  2003), resulting in  a fraction of 10$\%$
in  the innermost  field, 20$\%$  in the  intermediate one  and  up to
50$\%$  in the  outer one.  This percentage,  especially in  the outer
field, cannot be neglected and would obviously change the slope of the
gradient.  However  it cannot erase  the {\it presence} of  a gradient
(see Zoccali et al. 2008 for a discussion).

The bulge MDF  in Baade's Window obtained by Zoccali  et al. (2008) is
centered at the same mean metallicity than the one by Fulbright et al.
(2006).   However,  the  former  is significantly  narrower  than  the
latter,  likely  due to  the  smaller  errors  on the  metallicity  of
individual stars.  Recently Rangwala \& Williams derived the bulge MDF
in  a few  fields including  Baade's Window,  by means  of Fabry-Perot
photometry across  the Ca II  triplet line at  8542 \AA. Their  MDF is
compatible  with both  the Fullbright  et  al. (2006)  and Zoccali  et
al. (2008) results, once the  latter two are convolved with the larger
error of the Fabry-Perot measurements.

The presence  of a radial  metallicity gradient along the  bulge minor
axis favors a bulge  formation through dissipational collapse, against
the bar-driven  scenario. In fact,  the latter is a  dynamical process
that   should  not,  in   principle,  segregate   metallicities.   The
independent evidence of the absence  of a radial gradient in the inner
region ($<600$ pc)  cannot be contradicted here, thus  one should keep
in  mind  the  possibility  of  a  two  component  bulge,  uniform  in
metallicity  in   its  inner   part,  and  with   radially  decreasing
metallicity in the outer part.

Further evidence in the direction  of a two component bulge comes from
a detailed analysis  of the MDF in Baade's  Window.  The observed MDF,
once deconvolved  from the  estimated errors, appears  clearly bimodal
(Hill et  al.  2009).  Furthermore, the two  components seems  to have
different kinematics (see below).

It  is  worth mentioning  that  in the  past  few  years several  high
resolution spectroscopic analysis of bulge dwarf stars were carried on
during   microlensing  events   that   temporarily  magnifyied   their
brightness  (Johnson et  al.  2007,  2008;  Cohen et  al. 2008,  2009;
Bensby  et al.   2009a, 2009b).   The results  of these  analysis were
extremely surprising,  because the  mean [Fe/H] of  microlensed dwarfs
was too high to be compatible with a random (if small) sampling of the
bulge MDF obtained from giants.  Newer results, however, now including
a total of  13 bulge microlensed dwarfs, demonstrate  that the initial
discrepancy has almost completely  disappeared, and it was very likely
due  to   small  number  statistics  (Bensby  et   al.   2009c,  these
proceedings).

\section{Bulge element ratios}

Element  ratios  carry   important  information  about  the  formation
timescale  of a  stellar system.   In particular,  the ratio  of alpha
elements over iron  is a measure of the  relative contribution of type
II SNe  (producing mainly alphas)  relative to type Ia  SNe (producing
mainly iron). Therefore  stars with [$\alpha$/Fe] significantly higher
than 0, such  as the halo stars, were born before  the lower mass type
Ia SNe started  to explode.  McWilliam \& Rich  (1994) first suggested
that bulge  stars have alpha  element enhancement with respect  to the
Sun, suggesting a rapid star  formation for the bulge. Rich \& Origlia
(2005) confirmed  this result with  near IR  spectra, though  only for
stars in  a narrow range  of $-0.35<$[Fe/H]$<0$. Zoccali et  al (2006)
and Lecureur et al.  (2007) extended the former studies to a sample of
50 stars, observed with  UVES (R=45,000) simultaneously to the GIRAFFE
observations  mentioned   above,  and  spanning  $-0.8<$[Fe/H]$<+0.4$.
Their results confirmed the  alpha element enhancement of bulge stars,
well reproduced  by chemical evolution models assuming  a rapid ($\sim
1$  Gyr) star  formation timescale  (Immeli  et al.  2004; Ballero  et
al. 2007).

Lecureur et al.  (2007) found that different elements,  such as oxygen
and magnesium,  behave differently, when plotted  against [Fe/H], thus
supporting  theoretical  models  with  metallicity  dependent  stellar
yields  (c.f., McWilliam  et al.   2008; Cescutti  et al.   2009).  

By comparing the [$\alpha$/Fe] trend  of bulge K giant stars with that
of solar neighborhood  dwarfs (Bensby et al. 2004;  Reddy et al. 2006)
Zoccali et al.  (2006) and  Lecureur et al.  (2007) concluded that the
bulge is chemically  different from the (local) thin  and thick disks,
and it must have formed more rapidly than both of them. These findings
were confirmed by  Cunha \& Smith (2006), Fulbright  et al. (2007) and
Rich  et  al  (2007a).    More  recently,  Mel\'endez  et  al.   (2008)
questioned  the   above  conclusions.   By  means   of  a  homogeneous
comparison of new near IR spectra  of bulge K giants with similar data
for  thin and  thick disk  K giants,  they found  a similarity  in the
[O/Fe] abundances of bulge and  thick disk stars, supporting a similar
origin -or at least formation timescale- for both components.

The origin of the different result  of Mel\'endez et al. (2008) is the
systematically higher [O/Fe]  they found for both thin  and thick disk
giants, while the  bulge abundance ratios are consistent  with all the
previous  results.   They  correctly  emphasize the  importance  of  a
homogeneus comparison between the same kind of stars (K giants) in the
bulge, thin and  thick disk.  Some of the  previous studies (Fulbright
et al. 2007; Rich et al.  2005, 2007a) also included spectra for disk K
giants, analysed  in the same way  as the bulge ones,  and yet yielded
lower [O/Fe] than bulge  stars.  However, the separation between thick
and thin  disk stars for  this purposes must  be done on the  basis of
kinematics  only, it is  very tricky,  and it  has not  been discussed
extensively in  the papers mentioned above.   Further investigation is
needed in  order to clarify  whether the bulge  and the thick  disk do
share the same  element ratios, thus might have  similar origin (e.g.,
Alves Brito et al. 2009, these proceedings).

The new,  high resolution  (R=70,000) near IR  spectrograph CRIRES@VLT
allowed Ryde  et al.  (2009)  to obtain precise measurements  of C,N,O
elements in a sample of bulge K giants all included in the FLAMES-UVES
sample observed  by Lecureur et  al.  (2007).  Their [O/Fe]  for bulge
stars are compatible with most previous measurements, but have smaller
statistical errors. The comparison with the thick disk giants observed
by Mel\'endez et al.  (2008) confirmed the chemical similarity between
the two components. 

It might be worth mentioning that high resolution near IR spectra have
recently been obtained for a  sample of red supergiant in the Galactic
center ($<50$  pc).  All  the available studies  (Cunha et  al.  2007;
Davies  et  al.   2009;  Najarro  et  al.  2009)  agree  on  a  [Fe/H]
distribution sharply  peaked around $\sim  0.1$; and on  [O/Fe] ratios
evenly spread  between +0.05 and  +0.45. The reason for  alpha element
enhancement in this case  is rather unclear.  The independent evidence
of  intense star  formation  in this  region  (e.g., An  et al.  2009)
implies recent  formation of massive  stars -- thus core  collapse SNe
shortly  after --  naturally  enriching the  inter  stellar medium  of
oxygen  and other alphas.   Alternatively --  or simultaneously  -- it
might also be that the recent  star forming activity in the center was
fueled  by  gas  coming  from  the bulge/bar  (known  to  produce  gas
inflows), hence already enriched by  different kind of SNe and stellar
winds, explaining the unusual element ratios.

\section{The bulge kinematics}

Several bulge proper motions studies have been carried on in different
bulge  windows, mostly  for bright  stars.  Only  in a  few  cases the
photometry was deep enough to allow the kinematical decontamination of
the turnoff region of the CMD  ( Zoccali et al.  2001; Kuijken \& Rich
2002;   Clarkson  et  al.    2008).   Other   studies  aimed   at  the
characterization  of the  bulge rotation  and velocity  dispersion, in
order to  understand if  the bulge exibits  a solid body  rotation, if
there are streaming motions, asymmetries, and if there is any evidence
of different sub-populations  with different kinematics (Spaenhauer et
al. 1992;  Alcock et al.   2001; Sumi et  al. 2004; Rattenbury  et al.
2007a,b; Vieira et  al. 2007; Soto, Kuijken \&  Rich 2007).  Vieira et
al.   (2007,  their  Table~2)  give  a compilation  of  the  available
determinations  of proper motions  dispersions, all  compatible within
the errors  with their $\sigma(\mu_l)  cos b=3.39\pm 0.11$  mas/yr and
$\sigma(\mu_b)=2.91\pm 0.09$ mas/yr.

Stellar kinematics  can help understanding the nature  of the Galactic
bulge.   In fact,  it is  expected  that a  {\it pseudobulges}  formed
through  secular evolution  of the  disk will  have a  larger rotation
compared to classical  ones, and the rotation velocity  is expected to
be constant  with galactic latitude.   In the so-called  Binney (1978)
diagram,  showing the  ratio  of the  maximum  rotation velocity  over
velocity   dispersion,   versus    the   asymmetry   parameter,   {\it
pseudobulges} are expected to  lie above classical ones.  The galactic
bulge, with  its $V_{\rm  max}/\sigma \sim 0.65$  (Rich et  al. 2007b;
Minniti  \& Zoccali 2007)  is consistent  with classical  spheroids in
external  galaxies. However, the  latest results  of the  BRAVA survey
(Howard et al.   2009) demonstrated that the rotation  velocity of the
bulge is  constant with  latitude (cilyndrical rotation),  as expected
for {\it  pseudobulges} and opposed to a  rotation velocity decreasing
outwards,  typical of classical  spheroids (Combes  et al.   1990; Fux
1997,  1999;  Zhao et  al.   1996;  Athanassoula  \& Misiriotis  2002;
Athanassoula 2005)

Once  more, the  nature of  the Galactic  bulge seems  consistent with
either a  classical or  a {\it pseudo}  bulge, depending on  the tools
used to probe it. A possible solution comes from the evidence that the
metal  rich and metal  poor component  have different  kinematics, the
first  one more  typical  of a  bar-like  structure, the  second of  a
dynamically hot system (Soto et al. 2007; Babusiaux et al. 2009).

\section{Conclusions}
The nature  of the  Galactic bulge is  somehow puzzling.   Its stellar
population is old ($10-12$ Gyr)  and it has a metallicity distribution
compatible  with  chemical  enrichment  models assuming  a  fast  star
formation.   The abundance  ratio  of alpha  elements  over iron  also
supports a  short star formation timescale, certainly  more rapid than
that of the thin disk, and  possibly more rapid than that of the thick
disk. The presence of a  radial metallicity gradient, at least outside
$\sim 600$ pc, favors a formation scenario via dissipational collapse,
rather than secular evolution of the disk. Nevertheless, the bulge has
the  shape of a  bar (perhaps  including some  X-shape feature  in the
outer part) and a  cylindrical rotation velocity, both characteristics
of  a  {\it  pseudobulge}  formed  via dynamical  heating  of  a  bar,
resulting from disk secular evolution.

A possible  solution to  these conflicting results  may come  from the
confirmation  of a  double component  bulge, as  already  suggested by
several studies (e.g., Soto et  al. 2007, Hill et al.  2009, Babusiaux
et al. 2009) and seen in several bulges of external galaxies (Peletier
et al. 2007).

In any case,  it is now clear from  several independent evidences that
the Galactic  bulge is  a complex structure.   There is  a metallicity
gradient in the outer region that seems not to be present in the inner
region.   There are  indications that  the  MDF in  Baade's Window  is
bimodal, with  each of the two component  having different kinematics.
Stellar  ages are  predicted to  be  different along  the minor  axis,
compared to the edges of the  bar. Even the morphology itself does not
seem to be simply that of a bar, but rather something like an X-shape.
The properties of  the stellar population in Baade's  Window cannot be
considered  as   representative  of  the  whole   bulge:  larger  area
photometric  and  spectroscopic {\it  maps}  are  needed  in order  to
understand the  bulge structure and  origin.  The VVV survey,  and its
spectroscopic followups, will certainly reserve many surprises in this
sense.

\begin{acknowledgments}

This work was supported by the Fondap Center for Astrophysics 15010003,
CATA PFB-06, and Fondecyt Regular \#1085278.

\end{acknowledgments}

%_____________________________________________________________________

\end{document}